# Decreasing Utilization of Systems with Multi-Rate Cause-Effect Chains While Reducing End-to-End Latencies

Luiz Maia, Gerhard Fohler
University of Kaiserslautern-Landau, Germany
{maianeto,fohler}@rptu.de

*Abstract*—The Logical Execution Time (LET) model has deterministic properties which dramatically reduce the complexity of analyzing temporal requirements of multi-rate cause-effect chains. The configuration (length and position) of task's communication intervals directly define which task instances propagate data through the chain and affect end-to-end latencies. Since not all task instances propagate data through the chain, the execution of these instances wastes processing resources. By manipulating the configuration of communication intervals, it is possible to control which task instances are relevant for data propagation and end-to-end latencies. However, since tasks can belong to more than one cause-effect chain, the problem of configuring communication intervals becomes non-trivial given the large number of possible configurations. In this paper, we present a method to decrease the waste of processing resources while reducing end-to-end latencies. We use a search algorithm to analyze different communication interval configurations and find the combination that best decrease system utilization while reducing end-to-end latencies. By controlling data propagation by means of precedence constraints, our method modifies communication intervals and controls which task instances affect end-to-end latencies. Despite the sporadic release time of some task instances during the analysis, our method transforms those instances into periodic tasks. We evaluate our work using synthetic task sets and the automotive benchmark proposed by BOSCH for the WATERS industrial challenge.

## I. INTRODUCTION

In modern automotive systems, devices such as sensors and actuators have to constantly communicate in order to achieve some system functionalities. Normally, the control applications present in those systems are very sensitive to end-to-end (E2E) communication latencies. For them, the E2E latencies have to be within a given range, or the application's control performance can be compromised [1].

A typical example of a controlled application in the automotive domain is a sensor to actuator application. It consists of a task that reads the sensor (*cause*), a task that processes the read value, and a task that writes to an actuator (*effect*). This chained sequence of tasks executing and propagating data is known as a *cause-effect chain* (CEC). The analysis of whether or not the E2E latency between the *cause* and the *effect* fulfills required timing constraints is not trivial [2]. The complexity increases when the CEC contains tasks with different periods (multi-rate CEC) and multiple data dependencies. The complexity further increases when tasks can be mapped to different cores and execute in parallel, e.g., in multi-core systems [3].

The Logical Execution Time (LET) model [4] emerged as a method to reduce the complexity of analyzing the temporal requirements of multi-rate CECs. By having fixed inter-task communication points, LET adds timing and data flow determinism to the analysis of multi-rate CECs. In LET, inter-task communication only occurs at the boundaries of the so-called *communication interval* [5], which is considered to be equal to the period interval of the task. As a result, LET abstracts from the actual system implementation (scheduling choices), and consequently reduces analysis complexity, but at the cost of increased pessimism, i.e., larger E2E latency values.

Recently, Maia et al. [6] proposed a method to reduce the pessimism present in the LET model by shortening and shifting tasks' communication intervals. Additionally, Biondi et al. [7] observed that in a multi-rate CEC applying the LET model, not all task instances contribute to data propagation, meaning that their execution wastes processing resources.

In LET, the configuration (length and position) of the communication intervals defines if a task instance propagates data through the chain or not. Therefore, by manipulating the configuration of communication intervals, it is possible to control which task instances are relevant for data propagation and, consequently, E2E latencies. Since control applications can have multiple CECs and tasks can belong to more than one CEC, the problem of configuring communication intervals becomes a non-trivial task given the large number of possible configurations.

In this paper, we present a method to decrease system utilization while reducing E2E latencies of systems with multi-rate CECs applying the LET model. Based on the configuration of the communication intervals, our method identifies which task instances can have their execution skipped as they do not propagate data through the complete chain. By adding precedence constraints between specific task instances, our method further manipulates the communication intervals previously proposed by Maia et al. [6] obtaining a wider variety of interval configurations. Our method uses a search algorithm to analyze different communication interval configurations and find the combination that best fits our goal. Although skipping task instances to reduce system utilization may cause tasks to become sporadic, our method solves this issue by transforming the resulting task instances into periodic tasks.

During later design phases, our method can be applied to optimize system utilization and E2E latencies. That is, energy can be saved, background services can be enabled during runtime and actuators can react faster to new sensor events. If needed, our method can be applied individually to selected tasks and/or CECs, e.g., for legacy reasons. Our method also allows designers to favor reduced system utilization or E2E latencies. Evaluation results are based on the automotive benchmark presented by BOSCH [8] and synthetic task sets.

Summary of contributions: Our method

- decreases system utilization while reducing E2E latencies
- manipulates the configuration of communication intervals in order to control data propagation through the chain
- keeps tasks periodic and with well-defined periodic inter-task communication points
- can be applied individually to selected tasks and/or CECs for legacy reasons

## II. RELATED WORK

The two most commonly considered latencies when analyzing a multi-rate CEC are: *reaction time* (*First to First* semantic) and *data age* (*Last to Last* semantic) [9]. The reaction time measures the *reactivity* of the system. It is the time interval between the occurrence of an external event until the *first* output based on that event. Data age measures the *freshness* of data. It is the time interval between a data sampled (read) by the first task in the CEC until the *last* output (actuation) based on such data is produced by the last task in the CEC. Recently, Günzel et al. [10] showed that the values for the maximum reaction time (MRT) and maximum data age (MDA) are equivalent.

In the literature, Becker et al. [11] proposed to use job-level dependency as a way to control data propagation and E2E latencies in multi-rate CECs applying the implicit communication model. In [12], Dürr et al. introduced the concept of *job chains* and provided an analysis of the maximum reaction time and maximum data age. Schlatow et al. presented in [13] an analysis of the data age for periodic offset-synchronized tasks. In [14], Günzel et al. presented a timing analysis of asynchronized distributed CECs.

The LET model [4] was first introduced as part of the Giotto programming language in the context of time-triggered tasks. In [15], Biondi et al. presented a method to implement the LET model using additional dedicated tasks to realize the logical behavior of LET. Biondi et al. presented in [7] an implementation of the LET model on actual multi-core platforms for automotive systems. In [16], Pazzaglia et al. used LET to enforce causality and determinism as a way to control accesses to shared memory and optimize the functional deployment on multi-core platforms.

Techniques have been proposed to compute the E2E latencies of multi-rate CECs applying the LET model. In [17], Becker et al. presented a method to compute the maximum data age considering different communication models. Kordon and Tang [18] proposed a method to determine the maximum data age based on a task dependency graph. In [19], Martinez et al. presented a phase-aware LET analysis to improve the E2E latencies of multi-rate CECs. In [20], Bradatsch et al. proposed a method to reduce data age by setting the communication intervals equal to tasks' worst-case response time. Maia et al. [6] proposed a method that shortens and shifts the communication intervals of tasks applying the LET model. Wang et al. [21] presented a method that also manipulates the length and position of the communication intervals but considers constrained instead of implicit deadlines. In [22], Zhang et al. proposed a loosened LET model. Their method does not reduce E2E latencies when reducing utilization and the communication intervals of merged tasks are not periodic and do not have the same length.

Considering the observations made by Biondi et al. [7], we model our problem of reducing system utilization as a search tree. The goal is to find the set of communication intervals that best decreases system utilization while reducing E2E latencies. By using the idea of job-level dependencies presented by Becker et al. [11], we further manipulate the communication intervals proposed by Maia et al. [6] while keeping tasks periodic and with well-defined periodic inter-task communication points.

## III. SYSTEM MODEL

We consider a multi-core system and a task set $\Gamma$ containing periodic real-time tasks. Each task is pre-allocated to a specific core and $\gamma$ represents the set of tasks allocated on a given core.

### A. Tasks and Jobs

A task $\tau$ is a tuple $(C_\tau, T_\tau, D_\tau, \phi_\tau)$, where $C_\tau$ represents the worst-case execution time (WCET), $T_\tau$ is the period, $D_\tau$ is the deadline, and $\phi_\tau$ is the phase. We assume tasks have implicit deadlines, i.e., deadline is equal to period. A *job* $J$ represents an instance of $\tau$, where $J(i)$ is the $i^{th}$ instance of $\tau$, $i \in \mathbb{N}^+$. $J(i)$ has a release time at $\phi_\tau + (i-1)T_\tau$ and an absolute deadline $D_\tau$ time units later. A schedule $\mathcal{S}$ specifies the execution behavior of all jobs of $\tau$ according to fixed-priority scheduling policy. That is, each task $\tau \in \Gamma$ has a given priority level $P_\tau$, where $P_\tau < P_{\tau'}$ means that $\tau$ has lower priority than $\tau'$. The *start time* of $J(i)$ according to $\mathcal{S}$ is $s^{\mathcal{S}}_{J(i)}$, while the *finishing time* is $f^{\mathcal{S}}_{J(i)}$. If the choice of a schedule is clear, we omit the index $\mathcal{S}$ for all definitions. Since tasks might have phase, the invocation pattern of tasks in a given core repeats every $H(\gamma)$ after $\Phi(\gamma)$, where $\Phi(\gamma) = \max(\phi_\tau | \tau \in \gamma)$ and $H(\gamma) = LCM(T_\tau | \tau \in \gamma)$.

### B. Communication Model

We assume communication between tasks happens through the use of shared resources and to be based on the LET model. Each task $\tau$ has a fixed and well defined communication interval $L_\tau$ with length and position defined according to the method proposed by Maia et al. [6]. The inputs and outputs of $\tau$ are logically updated at the boundaries of $L_\tau$. $begin(L_\tau)$ and $end(L_\tau)$ are relative points in time w.r.t. the release time of $\tau$, they represent the boundaries of $L_\tau$, i.e., $L_\tau = [begin(L_\tau), end(L_\tau)]$. We represent the length of interval $L_\tau$ as $|L_\tau|$.

Each job $J$ of $\tau$ has a communication interval $L_J$. The boundaries of $L_J$ define when a job $J$ *logically* receives (reads) input from a shared resource, as well as when it *logically* transmits (writes) output to a shared resource. At $begin(L_J)$, the logical *read-event* of $J$ from a shared resource occurs. For instance, if $begin(L_\tau) = 0$, that means the logical read-event of each $J \in \tau$ happens during its release. At $end(L_J)$, the logical *write-event* of $J$ to a shared resource occurs. For instance, if $end(L_\tau) = T_\tau$, that means the logical write-event of each $J \in \tau$ happens at the end of its period.

Figure 1 shows the communication boundaries of interval $L_{J(i)}$ that has been shifted ($\phi_\tau$) and shortened for a given job $J(i)$, i.e., $|L_\tau| \neq T_\tau$. In Section IV, we do a brief recap on how communication intervals can be shortened and shifted.

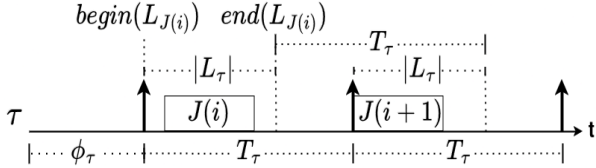

Fig. 1: Communication boundaries of interval $L_{J(i)}$ for a given job $J(i)$ of task $\tau$

### C. Cause-Effect Chain

A Cause-Effect Chain (CEC) represents an ordered sequence of communications carried out between a finite set of tasks. We represent a CEC by $E = (\tau_1 \to \tau_2 \to \cdots \to \tau_{|E|})$, $|E|$ being the number of tasks in $E$. The function $E(i)$ returns the $i^{th}$ task in $E$, $i \in \{1, 2, \cdots, |E|\}$. For a pair of tasks in $E$, the $\to$ operator indicates that $\tau_{i+1}$ acts as a consumer/reader task, while $\tau_i$ as a producer/writer task.

We assume that $E$ samples (acquires data) at every $begin(J_1)$, $J_1$ being a job of task $E(1)$. Like Günzel et al. [10], when computing MRT, we consider the maximum time interval between the occurrence of an external event (input) and its sampling by $J_1$. Likewise, when computing MDA, we consider the maximum time interval between $end(J_{|E|})$ and the actuation (output).

### D. Job Chain

Given a CEC $E$, a job chain $c^E = (J_1 \to \cdots \to J_{|E|})$ is a finite sequence of jobs representing one of the possible data propagation paths of $E$. In a job chain, the following requirements are respected:
- $J_i$ is a job of $E(i)$, $i \in \{1, 2, \cdots, |E|\}$.
- The data written by $J_i$ is read by $J_{i+1}$. That is, $end(L_{J_i}) \leq begin(L_{J_{i+1}})$ for all $i \in \{1, 2, \cdots, |E|-1\}$

Since in the LET model read and write events occur periodically according to communication intervals, every data propagation path of $E$ repeats after an interval of time. After $\Phi(E) + H(E)$, a data propagation path repeats every $H(E)$ time units, where $\Phi(E) = \max(\phi_{\tau_1}, \cdots, \phi_{\tau_{|E|}})$ and $H(E) = LCM(T_{\tau_1}, \cdots, T_{\tau_{|E|}})$. Therefore, for the sake of computing the E2E latencies of a CEC $E$, it is enough to analyze the job chains with external activity within one of the repetition intervals of $E$, e.g., $[\Phi(E)+H(E), \Phi(E)+2H(E))$.

Our notation is summarized in Table I.

| Variable | Definition |
|---|---|
| $\Gamma$ | task set under analysis |
| $\tau \in \Gamma$ | a task |
| $J(i), i \in \mathbb{N}^+$ | $i^{th}$ job of a task $\tau$ |
| $\mathcal{S}$ | schedule according to a fixed-priority schedule |
| $s^{\mathcal{S}}_{J(i)}$ | start time of $J(i)$ according to $\mathcal{S}$ |
| $f^{\mathcal{S}}_{J(i)}$ | finishing time of $J(i)$ according to $\mathcal{S}$ |
| $H(\gamma)$ | hyperperiod of the tasks in $\gamma$ |
| $L_\tau$ | communication interval of task $\tau$ |
| $begin(L_\tau)$ | relative point in time representing the start of $L_\tau$ |
| $end(L_\tau)$ | relative point in time representing the end of $L_\tau$ |
| $L_J$ | communication interval of job $J$ |
| $begin(L_J)$ | logical read-event of $J$ |
| $end(L_J)$ | logical write-event of $J$ |
| $H(E)$ | hyperperiod of the cause-effect chain $E$ |
| $E(i)$ | the $i^{th}$ task in $E$, $i \in \{1, 2, \cdots, |E|\}$ |
| $J_i$ | a job of $E(i)$, $i \in \{1, 2, \cdots, |E|\}$ |

TABLE I: Notation Table

## IV. SHORTENING AND SHIFTING COMMUNICATION INTERVALS

In this section, we do a brief recap on how to derive new communication intervals for tasks applying the LET model. By exploiting information from a feasible schedule, the method proposed by Maia et al. [6] reduces the pessimism present in the LET model while maintaining its deterministic properties and the periodicity of tasks.

Instead of setting $|L_\tau| = T_\tau$, i.e., $L_\tau = [0, T_\tau]$ $\forall \tau \in \Gamma$, the method proposed by Maia et al. [6] derives new relative points in time for $begin(L_\tau)$ and $end(L_\tau)$. By repositioning the boundaries of $L_\tau$ and therefore the boundaries of $L_J$, the method postpones the logical read-event of $J$ and prepones the logical write-event of $J$.

### A. Defining Schedule-Aware Intervals

In order to make $L_\tau$ schedule-aware, $\forall \tau \in \Gamma$, the length and position of $L_\tau$ have to be adjusted according to a new time interval $I_\tau$, where the length of $I_\tau$ is $C_\tau \leq |I_\tau| \leq T_\tau$. $begin(I_\tau)$ and $end(I_\tau)$ delimit the boundaries of $I_\tau$, i.e., $I_\tau = [begin(I_\tau), end(I_\tau)]$.
The length and position of $I_\tau$ are derived according to schedule $\mathcal{S}$. As explained in Section III-A, $\mathcal{S}$ specifies the start time $s_J$ and the finishing time $f_J$ for all $J \in \tau$.
Below we present the definition of two terms, *relative start time* ($S^{\mathcal{S}}_J$) and *relative finishing time* ($F^{\mathcal{S}}_J$), which are later used to derive the communication boundaries for $I_\tau$.

*Definition 1:* **Relative Start Time (of a Job)**. Let $J(i)$ be the $i^{th}$ job of task $\tau$ in schedule $\mathcal{S}$. The relative start time ($S^{\mathcal{S}}_{J(i)}$) of a job is the start time of $J(i)$ minus its release time.

$$S^{\mathcal{S}}_{J(i)} = s^{\mathcal{S}}_{J(i)} - (\phi_\tau + (i-1)T_\tau), \text{where } i \in \mathbb{N}^+ \quad (1)$$

*Definition 2:* **Relative Finishing Time (of a Job)**. Let $J(i)$ be the $i^{th}$ job of task $\tau$ in schedule $\mathcal{S}$. The relative finishing time ($F^{\mathcal{S}}_{J(i)}$) of a job is the finishing time of $J(i)$ minus its release time.

$$F^{\mathcal{S}}_{J(i)} = f^{\mathcal{S}}_{J(i)} - (\phi_\tau + (i-1)T_\tau), \text{where } i \in \mathbb{N}^+ \quad (2)$$

Depending on when each $J$ executes between its release and deadline, the values for $S^{\mathcal{S}}_J$ and $F^{\mathcal{S}}_J$ can change for each $J \in \tau$ (one job may execute early during its period, while another job may execute later). In order to keep the timing and data-flow determinism of LET when setting $L_\tau = I_\tau$, it is necessary to ensure that all $J \in \tau$ have a common periodic communication interval, i.e., $\forall J \in \tau, \{S^{\mathcal{S}}_J, F^{\mathcal{S}}_J\} \in I_\tau$.

The communication boundaries for $I_\tau$ can be computed using the *earliest relative start time* ($ES^{\mathcal{S}}_\tau$) and the *latest relative finishing time* ($LF^{\mathcal{S}}_\tau$) of a task $\tau$ based on $\mathcal{S}$.

*Definition 3:* **Earliest Relative Start Time (of a Task)**. Let $\tau$ be a task in schedule $\mathcal{S}$. The earliest relative start time ($ES^{\mathcal{S}}_\tau$) of $\tau$ is the minimum relative start time among all jobs of $\tau$ in $\mathcal{S}$.

$$ES^{\mathcal{S}}_\tau = \min_{\forall J \in \tau} S^{\mathcal{S}}_J \quad (3)$$

*Definition 4:* **Latest Relative Finishing Time (of a Task)**. Let $\tau$ be a task in schedule $\mathcal{S}$. The latest relative finishing time ($LF^{\mathcal{S}}_\tau$) of $\tau$ is the maximum relative finishing time among all jobs of $\tau$ in $\mathcal{S}$.

$$LF^{\mathcal{S}}_\tau = \max_{\forall J \in \tau} F^{\mathcal{S}}_J \quad (4)$$

In order to exemplify definitions 1 to 4, let us consider the example described in the following paragraph.

(**Example 1:**) Consider a control application consisting of a CEC with three tasks, where $E = (\tau_1 \rightarrow \tau_2 \rightarrow \tau_3)$, $\tau_1(1,5,5,0)$, $\tau_2(1,3,3,0)$, $\tau_3(1,5,5,0)$, $P_{\tau_3} < P_{\tau_1} < P_{\tau_2}$. Figure 2 shows a schedule $\mathcal{S}$ for the tasks present in the control application. Let us analyze task $\tau_3$, which has three jobs in $\mathcal{S}$. Following definitions 1 and 2, job $J(i)$ has $S_{J(i)} = 2$ and $F_{J(i)} = 3$. Job $J(i+1)$ has $S_{J(i+1)} = 2$ and $F_{J(i+1)} = 3$, while $J(i+2)$ has $S_{J(i+2)} = 1$ and $F_{J(i+2)} = 2$. Following definitions 3 and 4, $\tau_3$ has $ES_{\tau_3} = 1$ and $LF_{\tau_3} = 3$. By doing the same to the other tasks in $\mathcal{S}$, their $ES_\tau$ and $LF_\tau$ values are: $ES_{\tau_1} = 0$ and $LF_{\tau_1} = 2$ for $\tau_1$; $ES_{\tau_2} = 0$ and $LF_{\tau_2} = 1$ for $\tau_2$.

Based on the values of $ES_\tau$ and $LF_\tau$ for a given task $\tau$, the boundaries of $I_\tau$ can be shifted and shortened respectively. For instance, $I_\tau$ is shifted by applying an additional phase to task $\tau$, i.e., $\phi_\tau = \phi'_\tau + ES_\tau$, where $\phi'_\tau$ is $\tau$'s initial phase. Since the boundaries of $I_\tau$ are relative points in time and $\tau$ was shifted according to $ES_\tau$, $begin(I_\tau) = 0$ and $end(I_\tau) = LF_\tau - ES_\tau$. Therefore, by setting $I_\tau = [begin(I_\tau), end(I_\tau)]$, and $L_\tau = I_\tau$, the communication interval $L_\tau$ of $\tau$ is shortened and shifted. Figure 3 shows the schedule-aware intervals of $\tau_3$. Note that $\tau_3$ was shifted by $\phi_{\tau_3}$ and the communication interval is now $L_{\tau_3} = I_{\tau_3} = [0,2]$ instead of $L_{\tau_3} = [0,T_{\tau_3}]$, i.e., $[0,5]$.

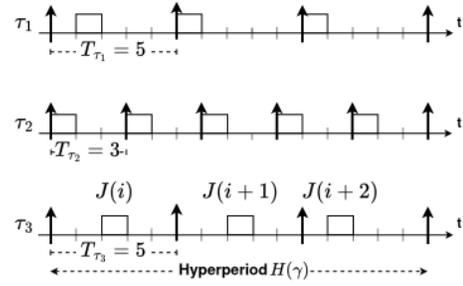

Fig. 2: Schedule $\mathcal{S}$ for a CEC $E = (\tau_1 \rightarrow \tau_2 \rightarrow \tau_3)$

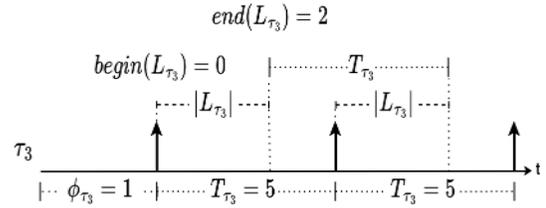

Fig. 3: New schedule-aware interval of task $\tau_3$

Note that although the method adds a phase $ES_\tau$ to a task $\tau$, neither the schedulability of the task set nor jobs' execution order in schedule $\mathcal{S}$ are affected: for any $J(i) \in \tau$, $i \in \mathbb{N}^+$, there is no $J(i)$ that executes before $(i-1)T_\tau + ES_\tau$ according to $\mathcal{S}$. All $J$ of $\tau$ have to wait at least $ES_\tau$ time units after its release in order to execute. Therefore, as long as the release of $\tau$ is postponed by $ES_\tau$ time units, $\forall \tau \in \Gamma$, the schedulability of the task set is not affected and the execution order of jobs in schedule $\mathcal{S}$ is preserved [6].

## V. IMPROVING SYSTEM UTILIZATION FOR SYSTEMS WITH MULTI-RATE CAUSE-EFFECT CHAINS

In this section, we present our method to reduce system utilization of control applications consisting of multi-rate CECs applying the LET model before runtime.

As explained by Gemlau et al. [23], in a producer-consumer relation between two tasks in a CEC applying the LET model, a job of the consumer task will always read (*consume*) from the same job of the producer task during any execution run of the CEC. Biondi et al. [7] observed that depending on the period of the tasks present in the CEC, not all jobs of a task need to update their shared resource. As a result, not all jobs of the tasks present in the CEC actively contribute to data propagation, i.e., those jobs do not affect the E2E latencies and their execution only wastes processing resources.

Although it may seem simple to recognize for a CEC $E$ all the data propagation paths ending within a repetition interval, i.e., hyperperiod $H(E)$, in a real-world example where several tasks with different periods are part of the same CEC, this turns out to be a non-trivial task. In Section V-A we show how to identify jobs that propagate data through complex CECs and how to decrease system utilization after identifying those jobs.

### A. Reducing System Utilization by Skipping Jobs

Given a set of CECs and the tasks composing them, our method determines for each CEC which jobs are responsible for data propagation by identifying the set of *primary job chains* ending within one hyperperiod $H(E)$. Below we provide the definition of the term *primary job chain*.

*Definition 5:* **Primary Job Chain**. Given a CEC $E$. Let $\Pi$ be a set containing job chains that end within one hyperperiod $H(E)$ and have the same job as their $J_1$. We call *primary job chain* $(pc^E)$, the job chain with the *earliest* $end(L_{J_{|E|}})$ in $\Pi$, i.e., with the lower $J_{|E|}$ index in $\Pi$.

The process of identifying for a CEC $E$ the primary jobs chains that are ending within one of the repetition intervals of $E$, e.g., $[\Phi(E)+H(E), \Phi(E)+2H(E))$, starts with a job $J$ of $\tau_{|E|}$ within the interval under analysis. Starting from $\tau_{|E|}$, our method recursively computes all job chains ending within the interval using the algorithms proposed by Maia et al. [6]. Once our method identifies a job chain, it is added to *primaryChainSet* replacing any previously added chain that does not respect Definition 5. We summarize in Algorithm 1 our method to obtain primary job chains.

---

**Algorithm 1** Identifying Primary Job Chains

*taskVector*      ▷ Contains tasks present in $E$
*primaryChainSet* = ∅      ▷ Contains primary job chains
$J_1$ *instances* = ∅      ▷ Contains $J_1$'s of the primary job chains
*jobVector*      ▷ Contains jobs of $\tau_{|E|}$ within one repetition interval.

1: **procedure** $identifyPrimaryChains(jobVector, taskVector)$
2:   **for** $job \in jobVector$ **do**
3:     $jobChain$ = findJobChain($job, taskVector$) ▷ Find job chains according to [6]
4:     **if** $J_1$ of $jobChain \notin J_1 instances$ **then**
5:       insert($jobChain, primaryChainSet$) ▷ Add $jobChain$ to *primaryChainSet*
6:       insert($J_1$ of $jobChain, J_1 instances$) ▷ Add $J_1$ to $J_1 instances$
7:     **else**   ▷ If *primaryChainSet* has a chain with $J_1$, check Definiton 5
8:       $storedChain$ = getStoredChain($J_1, primaryChainSet$)
9:       **if** $end(L_{J_{|E|}})$ of $jobChain < end(L_{J_{|E|}})$ of $storedChain$ **then**
10:         replaceChain($storedChain, jobChain, primaryChainSet$)
11:       **end if**
12:     **end if**
13:   **end for**
14: **end procedure**

---

In order to demonstrate how system utilization can be reduced after identifying primary job chains, let us consider again the control application and the tasks from Example 1 presented in Section IV. Figure 4 shows with *thick solid lines* the three primary job chains ending within one $H(E)$. Note that for a better representation of the complete propagation paths of the primary jobs chains ending within $H(E)$, as well as their MRT and MDA values, we have drawn a longer timeline. The job chains marked with *thin solid lines* are also primary job chains, but they belong to another repetition interval. For example, since job chains, repeat every $H(E)$, the chain starting at the fourth job of $\tau_1$ in Figure 4 is the same as the one starting at the first job of $\tau_1$ but one $H(E)$ apart. The job chains marked with *thin dotted lines* have their outputs overwritten by other job chains during data propagation. As a result, the execution of some jobs, e.g., $J(i)$, is only wasting processing resources. Note that in Figure 4 the communication intervals of the tasks were already shortened and shifted as in Section IV, i.e., $L_{\tau_1} = [0,2]$, $L_{\tau_2} = [0,1]$, $L_{\tau_3} = [0,2]$ and $\phi_{\tau_3} = 1$.

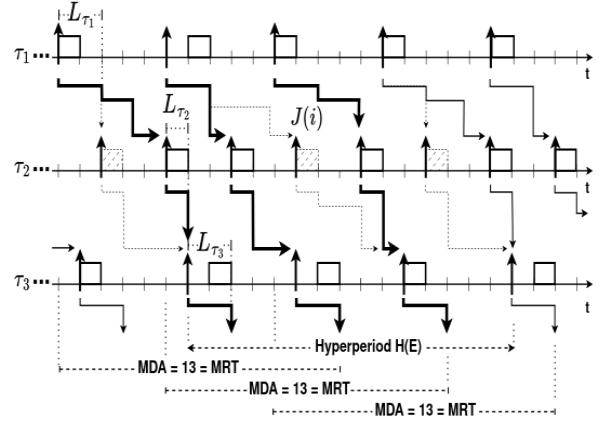

Fig. 4: Primary job chains of $E$ ending within $H(E)$

In Figure 4, we marked with dashed lines jobs that have their outputs overwritten. By skipping in each $H(E)$ the execution of jobs that are not part of primary job chains, system utilization can be reduced without affecting the E2E latencies of the control application. In a previous work, Maggio et. al [24] showed that control applications with timing requirements can skip the execution of some jobs without affecting the quality of the controlled application. Therefore, for control applications with CECs applying the LET model, it is safe to skip in each $H(E)$ the execution of jobs that are not part of primary job chains given the data flow determinism present in LET. For example, by skipping the execution of the dashed jobs of $\tau_2$ in Figure 4, the system utilization reduces from 0.73 to 0.6. Note that in order to not change the input/output behavior of the controlled application, jobs from tasks located at the boundaries of the chain are not to be skipped. Also, if a task belongs to more than one CEC, only jobs that are not needed in all those CECs are skipped.

In order to reduce system utilization while reducing E2E latencies, our method needs to manipulate the configuration of communication intervals (length and position) since they determine which jobs are part or not of primary job chains. In Section V-B, we show how to further manipulate communication intervals and how they can be used to reduce system utilization and E2E latencies.

## B. Manipulating Communication Intervals

In [6], Maia et al. showed that by manipulating the configuration of communication intervals, it is possible to control which jobs propagate data on multi-rate CECs applying the LET model. The proposed method derives new boundaries for the communication intervals by exploiting information from a feasible schedule. However, by deriving the intervals in this manner, their method becomes limited to only one configuration per schedule.

In order to decrease utilization while reducing E2E latencies, a method that is not limited to a specific set of communication intervals is needed. One way to obtain different interval configurations is by changing the order in which jobs execute. We consider that when a schedulable task set is available, it is possible to change the order in which some jobs execute by setting additional job-level dependencies (JLDs). We define a dependency ($\prec$) between two jobs as one can only start its execution once the other has finished it. For example, given a pair of jobs from two different tasks, if there is a JLD between them, the job holding the dependency has to wait for the other job's execution despite its priority or release time.

In the literature, JLDs were previously used by Becker et. al [11] to control data propagation through multi-rate CECs using communication models different than LET, e.g., the implicit communication model. In this work, we use JLDs to modify the boundaries of communication intervals in the LET model. By adding JLDs between specific jobs in a multi-rate CEC applying the LET model, our method can further manipulate the configuration (length and position) of communication intervals ($L_\tau$). As a result, new primary job chains are obtained and this ultimately affects which jobs in the CEC affect or not the E2E latencies. Therefore, depending on which and how many jobs are not part of primary job chains, system utilization can be further minimized by skipping the execution of those jobs. By testing different configurations for the communication intervals, we can select the configuration that best decreases system utilization while reducing E2E latencies.

Let us consider again Example 1 in order to demonstrate how JLDs can impact the configuration of communication intervals. Figure 5 shows that by introducing JLDs between specific jobs of $\tau_2$ and $\tau_3$, new communication intervals can be obtained.

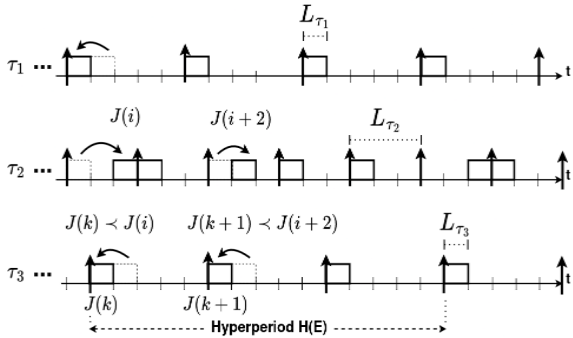

Fig. 5: $L_\tau$ after adding two JLDs between $\tau_2$ and $\tau_3$

Note that the curved arrows in Figure 5 are used to indicate the change in the execution order of the jobs. By adding one JLD between the first jobs of $\tau_2$ and $\tau_3$, i.e., $J(k) \prec J(i)$, the first jobs of $\tau_1$ and $\tau_3$ have their execution preponed, while the first job of $\tau_2$ has its execution postponed to the end of its period. By adding another JLD between the third job of $\tau_2$ and the second job of $\tau_3$, i.e., $J(k+1) \prec J(i+2)$, the third job of $\tau_2$ had its execution postponed, while the second job of $\tau_3$ had its execution preponed. Using the method described in Section IV, we obtain new communication intervals for all tasks, i.e., $L_{\tau_1} = [0, 1]$, $L_{\tau_2} = [0, 3]$, $L_{\tau_3} = [0, 1]$. Having new intervals lead to a new set of primary job chains, allowing our method to skip the execution of a different set of jobs, possibly resulting in a further reduction in system utilization. Note that a JLD added between two jobs in a repetition interval is also added in all the other intervals.

As long as the task set remains schedulable, $n$ additional JLDs can be added to the task set, $n \in \mathbb{N}$. Therefore, it is possible to define different configurations for $L_\tau$ depending on how the execution order of jobs is arranged according to the additional JLDs. There are multiple options to find the configuration of $L_\tau$ which best decreases system utilization while also reducing E2E latencies. For instance, search algorithms or ILP solvers can be used to find the set of additional JLDs that best fits our objective. For convenience, in this paper we build our work on top of an existing scheduling framework [25], which contains an improved version of the PIDA* algorithm.

We model our problem of decreasing system utilization while reducing E2E latencies as a search tree. The root node represents a feasible schedule of task set $\Gamma$ according to fixed-priority scheduling policy. Algorithm 2 shows the creation of child nodes. In the following paragraphs we describe how Algorithm 2 constructs the search tree and how our heuristic works. Note that by means of precedence constraints, our algorithm tries to steer data propagation in a way that E2E latencies are reduced and the minimum number of jobs from tasks belonging to multiple CECs are needed.

*1) Search Algorithm:* In our search tree, every node represents a feasible schedule of $\Gamma$ after the addition of a JLD between two jobs. Starting from the root node, our algorithm selects the first job in the *jobVector* to receive a JLD ($job_X$) from jobs of other tasks (line 5). We call as *candidate job*, jobs that execute in between the release and deadline of $job_X$ (line 6). Our algorithm iterates over the list of possible candidates and checks if a feasible schedule can be obtained after a JLD is added between the selected job and one of the candidate jobs (lines 7 to 9). The algorithm generates a new node (*child node*) for every candidate job that keeps the task set schedulable after the JLD addition (line 10). For every child node, the algorithm recomputes the communication intervals of the tasks and checks the current values for system utilization and E2E latencies (lines 12 and 13).

*2) Search Heuristic:* Our heuristic guides the search on the tree by selecting the child node that best decreases system utilization while reducing E2E latencies. Depending on how the search tree is being traversed, i.e., which JLDs have been assigned on previous nodes, it is beneficial for the goal of our heuristic to consider the case where no JLD is assigned on the current node (line 17). Note that the generation of a child node containing no changes to the schedule is also important for the cases when none of the candidates jobs generated a feasible schedule after the JLD addition. By iterating over the possible candidates, our algorithm sorts child nodes according to their system utilization, MRT, and MDA values (line 18). The child node that best aligns with our searching goal is selected as the next node to be investigated (line 20).

Upon a time out defined by the user, our algorithm stops and returns the set of additional JLDs that best decreases system utilization and E2E latencies among the searched nodes.

**Algorithm 2** Creation of child nodes and sorting according to heuristic

    *cecVector*    ▷ Contains CECs sorted according to their length
    *jobVector*    ▷ Contains jobs sorted according to the *cecVector*
1: **procedure** $createAndSortChildNodes(node, childNodes)$
2:     **if** TIMEOUT **then**
3:         exit
4:     **else**
5:         $job_X = jobVector[node.nextJob]$ ▷ Get the next job to receive JLDs
6:         $candidatesSet = getCandidates(job_X)$ ▷ Get a set of jobs that can serve as JLDs for $job_X$
7:         **for** candidate $\in$ *candidatesSet* **do**
8:             $setJLD(job_X, candidate)$ ▷ Set JLD: candidate $\prec job_X$
9:             **if** isSchedulable($\Gamma$) **then** ▷ Checks $\Gamma$'s schedulability
10:                 $childNode = newChild(node)$ ▷ Create a new child node
11:                 $computeNewCommIntervals()$ ▷ Recompute $L_\tau, \forall \tau \in \Gamma$
12:                 $computeSystemUtilization(cecVector)$
13:                 $computeE2eLatencies(cecVector)$
14:                 $insert(childNodes, childNode)$ ▷ Add to *childNodes* vector
15:             **end if**
16:         **end for**
17:         $insert(childNodes, node)$▷ Child node without the addition of a JLD
18:         $heuristic(childNodes)$ ▷ Sort the child nodes according to system utilization, maximum reaction time (MRT) and data age (MDA)
19:     **end if**
20:     **return** ▷ PIDA* will select new nodes to be investigated based on the sorting order of the child nodes
21: **end procedure**

In Section VI, we demonstrate a possible solution on how to resolve complex constraints such as JLDs before runtime and how to keep task's releases periodic even when skipping jobs.

## VI. TRANSLATING SPORADIC JOBS INTO PERIODIC TASKS WHILE RESOLVING DEPENDENCY CONSTRAINTS

Due to the data flow determinism present in the LET model, system utilization can be decreased by skipping the execution of jobs that are not part of primary job chains. However, skipping job's execution in a periodic task results in a sporadic release of jobs for that same task. For example, let us consider in Figure 6 the tasks from Example 1 and the JLDs from Figure 5. Although $\tau_2$ executes five jobs within $H(E)$, the execution of $J(i)$ and $J(i+3)$ can be skipped to decrease system utilization as they are not part of primary job chains. In order to skip the execution of $J(i)$ and $J(i+3)$, $\tau_2$ should not release those two jobs when they were supposed to. By doing so, $\tau_2$ no longer behaves as a periodic task, but as a sporadic task. Note that in Figure 6 the JLDs between $\tau_2$ and $\tau_3$ resulted in new job chains that reduced the overall E2E latencies from 13 to 12 time units compared to Figure 4.

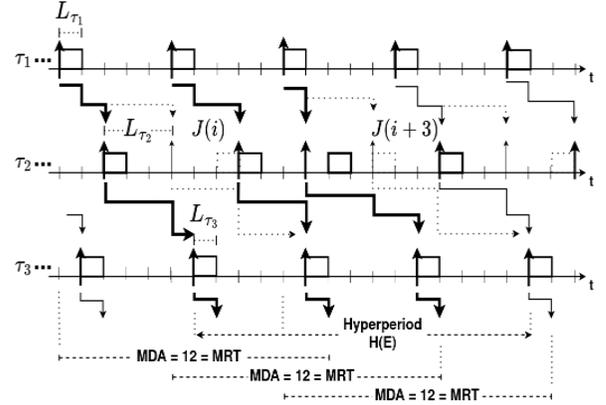

Fig. 6: Sporadic job release of $\tau_2$ due to job skipping

In order to obtain a sporadic release pattern but using a period task, we build our method on top of the work done by Dobrin et al. [26]. Their method translates offline schedules into fixed-priority schedule schemes while coping with complex timings requirements. The goal of our method is to translate into periodic tasks jobs that were not selected to be skipped by our algorithms. For example, Figure 7a shows in detail the sporadic release pattern of task $\tau_2$ from Figure 6. Figure 7b shows that the same execution pattern showed in Figure 7a can be obtained by translating $\tau_2$'s jobs into three periodic tasks. Note that $\tau_{2,3}$'s execution is delayed due to $\tau_3$.

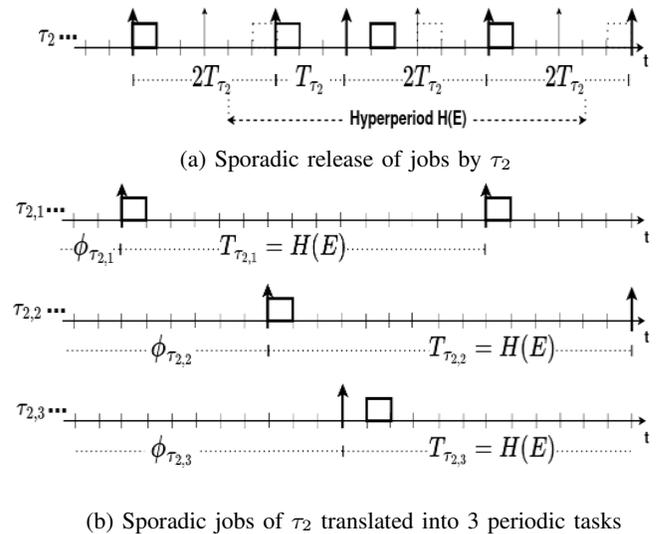

(a) Sporadic release of jobs by $\tau_2$

(b) Sporadic jobs of $\tau_2$ translated into 3 periodic tasks

Fig. 7: Translation of sporadic jobs into periodic tasks

By using task set $\Gamma$ and the outputs of algorithms 1 and 2 as input, our method transforms $\Gamma$ into $\Gamma'$ by adjusting task's parameters and translating specific jobs into periodic tasks. In order to generate $\Gamma'$, our method adds to a job set $\zeta$ each $J$ that is part of a primary job chain and for each one of them, our method checks if there are *offset assignment conflicts* and/or *priority assignment conflicts*. By resolving offset assignment conflicts, our method solves the sporadic releases of jobs when skipping jobs, while resolving priority assignment conflicts resolves JLDs before runtime. In the following paragraphs, we explain the two types of conflict and how to resolve them.

*1) Offset Assignment Conflict:* An offset assignment conflict exists if two consecutive jobs of a task $\tau$ in $\zeta$, require different offset values. That is, given two consecutive jobs of $\tau$ in $\zeta$, $J(i)$ and $J(k)$, where $i < k$, an offset assignment conflict exists if $k - i > 1$. For any two consecutive jobs of $\tau$ in $\zeta$ which $k - i = 1$ does not hold to be true, our method splits $\tau$ into $m$ instances, where $m$ is the number of jobs $\tau$ has in $\zeta$. Therefore, each $J$ of $\tau$ in $\zeta$ becomes a task $\tau_J$ that has $C_{\tau_J} = C_\tau$, $P_{\tau_J} = P_\tau$, $L_{\tau_J} = L_\tau$. The offset of $\tau_J$ is: $\phi_{\tau_J} = \phi'_\tau + (i-1)T_\tau - \lfloor \frac{S_{J(i)}}{H(E)} \rfloor H(E)$, where $i$ is the $i^{th}$ instance of $J$ and $\phi'_\tau$ is $\tau$'s initial phase. We define the period of $\tau_J$ as $T_{\tau_J} = H(E)$. As a result, each $\tau_J$ has only a single job within $H(E)$. During runtime, phase $\phi_{\tau_J}$ ensures that $\tau_J$ will start its execution at the same point in time as job $J$ would have started in the original schedule. Therefore, $\tau_J$ acts like an execution replacement for $J$. For example, the first job of $\tau_{2,3}$ in Figure 7b replaces the third job of $\tau_2$ in Figure 7a.

*2) Priority Assignment Conflict:* A priority assignment conflict exists when jobs of the same task in $\zeta$ require different priority levels to ensure a given execution order. That is, a priority assignment conflict exists for every $J$ in $\zeta$ that has a JLD. For each conflict $J' \prec J$, where $J$ is a job of $\tau$ and $J'$ a job of $\tau'$, our method splits $\tau$ into $m$ new tasks, where $m$ is the number of jobs $\tau$ has in $\zeta$. Job $J$ becomes a task $\tau_J$ with $\Psi < P_{\tau_J} < P_{\tau'}$, where $\Psi = \max(P_\tau | \tau \in \psi)$ and $\psi$ is the set of task with priority lower than $P_{\tau'}$. The other jobs of $\tau$ become tasks with the same priority as $\tau$. The remaining parameters of the new tasks are set in the same manner as when resolving offset assignment conflicts.

Figure 8 shows an example of the offset (resp. priority) assignment conflicts present in Figure 6 as a result of skipping jobs (resp. adding JLDs).

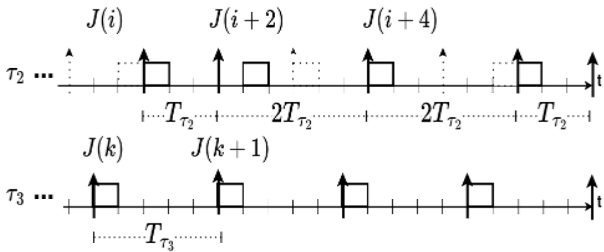

Fig. 8: Offset and priority assigment conflicts of $\tau_2$

In Figure 8, an offset assignment conflict between $J(i+2)$ and $J(i+4)$ exists because $J(i+3)$ is not part of a primary job chain (as shown in Figure 6). A priority assignment conflict between $J(i+1)$ and $J(i+2)$ exists due to the JLD added between $J(k+1)$ and $J(i+2)$. By resolving the offset (resp. priority) assignment conflicts of $\tau_2$ before runtime, our method obtains in Figure 9 the same execution pattern as in Figure 7a, but with reduced system utilization and without the need of JLDs between $\tau_2$ and $\tau_3$. Note that available solutions to enforce the execution behavior of JLDs during runtime can also be used [11] [27].

Figure 9 shows what the schedule previously showed in Figure 6 looks like after resolving the offset (resp. priority) assignment conflicts. Note that in Figure 9, our method translated $\tau_2$ into three periodic tasks $(\tau_{2,1}, \tau_{2,2}, \tau_{2,3})$. Although the parameters of $\tau_2$ changed, the E2E latencies of the control application (between $\tau_1$ and $\tau_3$) are the same as in Figure 6.

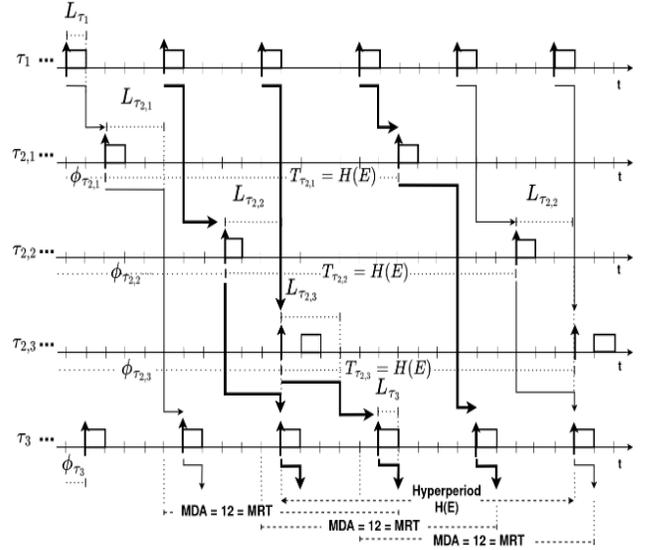

Fig. 9: Job chains after translating jobs into tasks

## VII. EXPERIMENTAL RESULTS

We evaluate our work based on the Real World Automotive Benchmarks presented by BOSCH [8] and synthetic task sets. With respect to the E2E latencies, we compare our results with the approach presented by Maia et.al [6] and the LET model. We consider that the task sets run on a system comprising four cores and that tasks are scheduled according to fixed-priority. For these set of experiments we tuned our heuristic to favor system utilization reduction over E2E latencies. We let our search algorithm run for 1 minute for each task set.

### A. Real World Automotive Benchmarks

We generated and tested 500 schedulable task sets based on the parameters of the Real World Automotive Benchmarks [8]. We assign periods to tasks following the definitions of Table III in [8]. The range of possible periods is: [1, 2, 5, 10,

20, 50, 100, 200, 1000]ms. Since, we do not consider angle-asynchronous tasks, we divided all probability values by 0.85. Inter-task communications follow the definitions of Table II in [8]. For each task we generated a WCET following the definitions of Tables IV and V in [8].

In our experiments, on average, there are 38 CECs per task set. The number of periods per CEC is randomly chosen between the interval [1,3] following the definitions of Table VI in [8]. For each period that composes the CEC, there are 2 to 5 tasks with that same period (Table VII in [8]). Each CEC is composed of 2 to 15 tasks. The total utilization of cores before optimization is $\approx 71\%$ (per core), on average.

Figure 10 shows that our method reduced system utilization by an average of $\approx 28\%$.

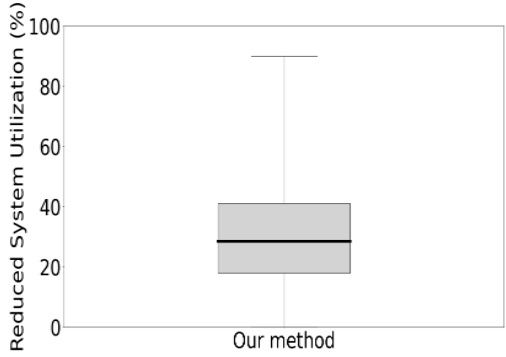

Fig. 10: Reduction in system utilization compared to LET

Since the values for the maximum reaction time and data age are equivalent (Günzel et al. [10]), we show the results for the maximum reaction time (MRT) in Figure 11. Note that in Figure 11, we normalized the results with respect to the maximum reaction time obtained by the LET model.

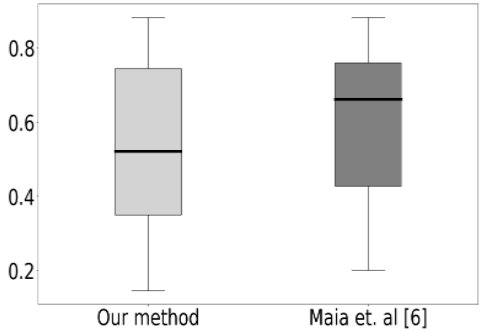

Fig. 11: Normalized maximum reaction time w.r.t LET

Figure 11 shows that while decreasing system utilization, our method reduced the MRT in $\approx 57\%$, on average, compared to the LET model. Compared to Maia et. al [6], our method further reduced the MRT in $\approx 14\%$, on average.

## B. Synthetically Generated Workloads

We randomly generated 500 schedulable task sets, where we chose task periods and inter-task communication as in the previous experiment. However, this time we allowed tasks to have higher WCET values, increased the number of possible periods per CEC from 3 to 5 and reduced the probability of single-rate CECs from 70% to 7%. The 63% difference was split equally among the other possible number of periods. The new probabilities for possible number of periods per CEC are {1: 7%, 2: 35.75%, 3: 25.75%, 4: 15.75%, 5: 15.75%}.

As a result of increasing the number of possible periods in a CEC, we increased the total number of tasks composing it from 15 to 25. We chose randomly the amount of CECs per task set from interval [10, 20], with an average of 13 CECs per task set. The total utilization of cores before optimization is $\approx 80\%$, on average. The obtained results are summarized in figures 12 and 13. Note that in Figure 13, we normalized the results with respect to the MRT obtained by the LET model.

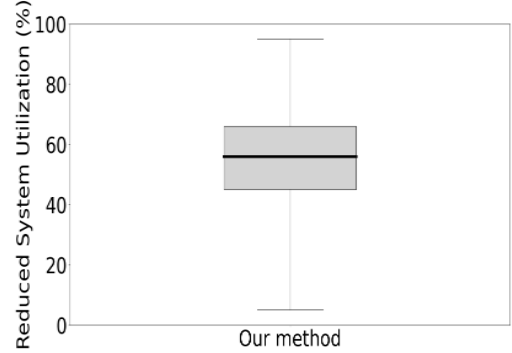

Fig. 12: Reduction in system utilization compared to LET

As shown in Figure 12, our method once again improved system utilization. For synthetic task sets, our model managed to reduce system utilization, in $\approx 55\%$, on average. Figure 13 shows that while manipulating communication intervals to decrease system utilization, our method reduced the MRT in $\approx 60\%$, on average, compared to the LET model, and $\approx 10\%$ compared to Maia et. al [6].

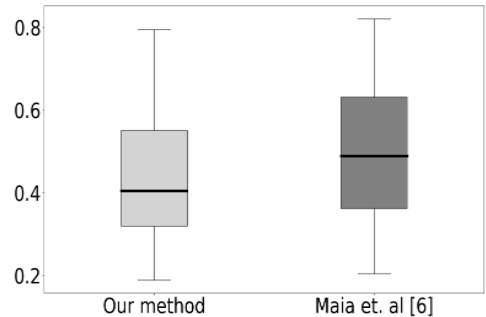

Fig. 13: Normalized maximum reaction time w.r.t LET

## VIII. Conclusion

In this paper, we proposed a method to decrease utilization and end-to-end latencies of systems with multi-rate cause-effect chains applying the LET model.

Our method uses a search algorithm to analyze different communication interval configurations and find the combination that best reduces system utilization and end-to-end latencies. By manipulating the configuration of communication intervals and skipping the execution of specific task instances, our method decreases system utilization while reducing end-to-end latencies. In order to maintain the deterministic characteristics of the LET model and tasks' periodicity, our method keeps well-defined periodic inter-task communication points and translates specific task instances into periodic tasks.

Experiments showed that for task sets based on the Real World Automotive Benchmarks presented by BOSCH [8] or synthetically generated, our method resulted in lower system utilization and end-to-end latencies. Considering all the evaluated task sets, our method reduced system utilization in $\approx 41\%$, on average, while the end-to-end latencies were reduced in $\approx 58\%$, on average, compared to the LET model. If needed, e.g., for legacy reasons, our method does not have to be applied to the entire task set, but also to only a subset. In future work, we plan to investigate the impact of our method in cause-effect chains containing different execution models and how it affects system utilization and end-to-end latencies.

## Acknowledgments

This work is supported by the Fonds National de la Recherche Luxembourg (FNR) and the Deutsche Forschungs Gemeinschaft (DFG) through the Core-Inter project ReSAC.